\documentclass[a4paper,11pt]{article}
\pdfoutput=1
\usepackage{graphicx}
\usepackage{epsf,amsmath,bbold,amsfonts,stmaryrd}

\usepackage[utf8]{inputenc}
\usepackage{mathrsfs}
\usepackage{appendix}
\usepackage{amssymb}
\usepackage{float}
\usepackage{color}
\usepackage{cite}
\usepackage{hyperref}
\hypersetup{pageanchor=false}
\usepackage{indentfirst}
\usepackage{url}
\usepackage{float}
\usepackage{caption}
\usepackage[numbers,square,comma,sort&compress,merge]{natbib}
\usepackage{graphicx}
\usepackage{animate}
\usepackage{subfigure}

\def\be{\begin{equation}}
\def\ee{\end{equation}}

\def\bea{\begin{eqnarray}}
\def\eea{\end{eqnarray}}

\def\ba{\begin{array}}
\def\ea{\end{array}}

\def\bc{\begin{center}}
\def\ec{\end{center}}

\def\bl{\begin{flushleft}}
\def\el{\end{flushleft}}

\def\br{\begin{flushright}}
\def\er{\end{flushright}}

\def\bi{\begin{itemize}}
\def\ei{\end{itemize}}

\def\bt{\begin{tabular}}
\def\et{\end{tabular}}

\numberwithin{equation}{section}

\begin{document}

\title{\textbf{Bending the Bruhat-Tits Tree II \\
the p-adic BTZ Black hole and Local Diffeomorephism on the Bruhat-Tits Tree }}
\author{Lin Chen, Xirong Liu \footnote{Chen and Liu  are co- first authors of the manuscript.}, and Ling-Yan Hung\\
   $^1$State Key Laboratory of Surface Physics, \\
    Fudan University, \\
    200433 Shanghai, China\\
    $^2$Shanghai Qi Zhi Institute, \\
    41st Floor, AI Tower, No. 701 Yunjin Road,  \\
    Xuhui District, Shanghai, 200232, China\\
    $^3$Department of Physics and Center for Field Theory and Particle Physics, \\
    Fudan University, \\
    200433 Shanghai, China\\
    $^4$Institute for Nanoelectronic devices and Quantum computing, \\
    Fudan University, \\
    200433 Shanghai , China\\}
\date{\today}

\date{}

\maketitle

\vspace{-10mm}

\vspace{8mm}

\begin{abstract}
In this sequel to \cite{paper2}, we take up a second approach in bending the Bruhat-Tits tree. Inspired by the BTZ black hole connection, we demonstrate that one can transplant it to the Bruhat-Tits tree, at the cost of defining a novel ``exponential function'' on the p-adic numbers that is hinted by the BT tree.
 We demonstrate that the PGL$(2,Q_p)$ Wilson lines \cite{Hung:2018mcn} evaluated on this analogue BTZ connection is indeed consistent with correlation functions of a CFT at finite temperatures. We demonstrate that these results match up with the tensor network reconstruction of the p-adic AdS/CFT with a different cutoff surface at the asymptotic boundary, and give explicit coordinate transformations that relate the analogue p-adic BTZ background and the ``pure'' Bruhat-Tits tree background. This is an interesting demonstration that despite the purported lack of descendents in p-adic CFTs, there exists non-trivial local Weyl transformations in the CFT corresponding to diffeomorphism in the Bruhat-Tits tree. 
\end{abstract}
\baselineskip 18pt

\thispagestyle{empty}
\newpage

\tableofcontents

\section{Introduction}
p-adic AdS/CFT as a toy version of the AdS/CFT was proposed in \cite{Gubser:2016guj, Heydeman:2016ldy}. 
An $n$-dimensional CFT  lives on a number field $Q_{p^n}$, and the purported dual bulk geometry whose isometry is the conformal group of the CFT
$SL(2,Q_{p^n})$ is a discrete graph called the Bruhat-Tits (BT) tree. 
The BT tree is thus the analogue of the pure AdS space.  As in the usual AdS/CFT dictionary where computing correlation functions of a small number of operator insertions 
can be done by considering quantum field theories in the pure AdS background without worrying about back-reaction to the geometry, the p-adic AdS/CFT works in the same way.
Witten diagrams on the BT tree have been studied extensively \cite{Gubser:2016guj, Chapman:2016hwi, Gubser:2017tsi, Qu:2018ned, Jepsen:2019svc, Garcia-Compean:2019jvk, Qu:2019tyi, Chapman:2016hwi}, and the results satisfy all the expected features of the p-adic CFT \cite{Melzer:1988he}. 
A natural question would be to ask whether one can include dynamics of the background so that the bulk theory becomes an analogue of gravitational theory on AdS. 
Some attempts have been made in \cite{Gubser:2016htz}, where each edge of the graph is assigned a dynamical length and that an action is proposed that governs the dynamics of the edge lengths. 
The action is the analogue of the Einstein Hilbert action, given by the sum of local graph curvatures. The matter action was also {\it covariantized} by coupling the matter fields to these edge lengths. 

This actually poses a puzzle. It is well known that the p-adic CFT does not accommodate descendants. This does not mean that the representations of the conformal group is finite dimensional -- in fact it is known in the Mathematics literature that all finite dimensional representations of the p-adic conformal group $SL(2,Q_p)$ is in fact trivial (i.e. one dimensional). See for example \cite{Ebert:2019src}.  It is just that Lie algebra is not well defined because the exponential map could take an exponent with finite p-adic norm to infinity.  Moreover, functions taking p-adic numbers to the reals are locally constant. Therefore usual derivatives of correlation functions vanish. It would thus appear that the stress tensor is not well-defined in the p-adic CFTs \cite{Melzer:1988he}. The stress tensor is however the central ingredient in the AdS/CFT correspondence that is identified with the bulk metric. 
Therefore it is not entirely clear what the edge dynamics correspond to from the perspective of the p-adic CFT. 

In our prequel \cite{paper2}, the tensor network reconstruction of the p-adic CFT partition function is shown to naturally recover a bulk action and a covariantized matter action. The edge lengths did not show up as an independent set of operators in the CFT, but rather as a Fisher information metric between states with various non-local operator insertion. This is not in any contradiction with known AdS/CFT dictionary, since the Ryu-Takayanagi formula would also imply that distances in the bulk are related to entanglement of the boundary CFT \cite{Ryu:2006bv}. 
Yet, for technical reasons we worked in the perturbative limit away from the {\it pure BT} background. It would be interesting to understand how more exotic and non-perturbative backgrounds such as black holes can be described in the BT bulk. 
Moreover we have not yet resolved the mystery about descendents, and relatedly, whether local conformal transformation and Weyl transformations are well defined in the CFT, and if they do, how they manifest themselves in the BT bulk. 

In this paper, we revisit the problem of the BTZ black hole in the BT background. It has been noted that similar to the AdS$_3$ case the BT tree can be orbifolded to produce structures that carry higher genus and they appear to be analogues of the handle-bodies studied in the AdS case. In particular, one can generate genus curves that appear to be analogues of the BTZ black holes \cite{Heydeman:2016ldy, Hung:2019zsk, Ebert:2019src}. 
We were initially inspired by the tensor network formulation that was shown to be reproduced by a Wilson line network in a purported $SL(2,Q_p)$ Chern-Simons gauge theory \cite{Hung:2018mcn}. We looked into Wilson lines evaluated on flat connections that mimic results obtained from the BTZ black hole formulated in terms of $SL(2,\mathbb{C})$ connections. The study however led us to observations that apply generally, independently of the tensor network construction. Our presentation will however start with generalities before zooming in on the Wilson line interpretation. 
What we would like to demonstrate in this paper is that the p-adic version of the BTZ black hole actually carries a way more intimate semblance with the AdS$_3$ sister. 
There exists a set of natural coordinates on the BTZ black hole that is the analogue of black hole coordinates in AdS$_3$. Moreover, there exists a non-trivial coordinate transformation that connects the black hole coordinates with the analogue {\it Poincare coordinates} advocated in \cite{Gubser:2016guj}. 
To do so, a novel notion of ``exponential'' is introduced. This would be the first example of a local Weyl transformation defined in the p-adic CFT that is explicitly demonstrated to be equivalent to bulk diffeomorphism in the BT tree.  These results are described in section 2. 

Then we would describe how the p-adic Chern-Simons formulation admits also the analogue of the BTZ $SL(2,Q_p)$ connection, and that correlation functions would be correctly reproduced again using the Wilson line network that is again in complete agreement with the tensor network covering the BT tree now with a different cutoff surface -- now the constant "$r$" surface in black hole coordinates.
Moreover, with the explicit form of the gauge connection written down, we are able to make comparison with AdS$_3$ case and identify concretely the black hole horizon, temperature and black hole entropy of the BTZ black hole.  These results are presented in section 3 and 4. 

We will conclude in section 5.

\section{Weyl transformation and diffeomorphism on the Bruhat-Tits tree}

In this section, we will describe a novel coordinates system that is most natural for the BT tree when it is arranged to describe the analogue BTZ black hole. 
We will also describe how this {\it black hole coordinates} can be related to the usual Poincare coordinates via a coordinate transformation analogous to the corresponding transformation in AdS$_3$ \cite{Maldacena:1998bw}.

\subsection{Poincare coordinates on the Bruhat-Tits Tree}

\begin{figure}[htbp!]
\centering

\includegraphics[width=0.45\textwidth]{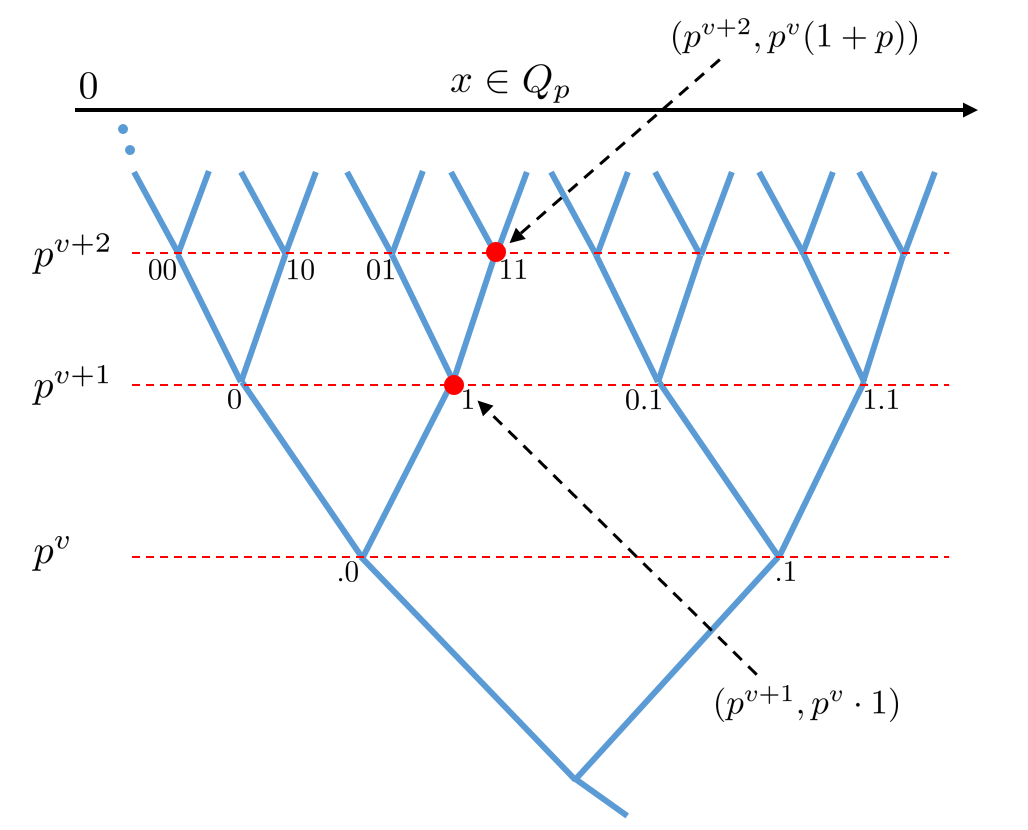}

\caption{The coordinates on the BT tree. }
\label{BTcoord}
\end{figure}

There is a full set of coordinates on the BT tree used in the mathematics literature. This is reviewed in detail in \cite{Bhattacharyya:2017aly,Hung:2018mcn} that would allow the action of the isometry on the tree completely explicit.  
 This  is the analogue of the Poincare coordinates. This coordinate system was taken and somewhat simplified in \cite{Gubser:2016guj} that hides certain detail. We will briefly review it here. Recall that the asymptotic boundary of the tree is the $Q_p$ line, and every boundary point has a unique path that connects it to a main branch, which encodes the p-adic number. There is a horizontal coordinate $x^{(n+1)}\in Q_p$ which is the analogue of $x$  in AdS poincare coordinates. The ``radial coordinate'' $p^{n+1}$ encodes the ``accuracy''  of $x^{(n+1)}$. i.e. $x^{(n+1)}$ is defined up to identification by adding to it an arbitrary $y$ with norm $|y|_p \le p^{-(n+1)}$. We might as well set all coefficients of  $p^{l+1}, \, l\ge n$ in $x^{(n+1)}$ to be zero.
Then the coordinates $(p^{n+1},x^{(n+1)})$ would take the following form
\begin{eqnarray}
  x^{(n+1)} &=& p^v\sum_{i=0}^{n-v} a_i p^i
\end{eqnarray}
as shown in Fig. \ref{BTcoord}.
It may appear somewhat unusual that the $x$ and radial coordinate $p^n$ appears somewhat intertwined. Also, the fact that it has two labels would make it appear very different from its AdS$_3$ sister. 

Let us massage this coordinate to make it look more familiar. 

\subsubsection{The $Q_p$ ``plane'' and its {\it three dimensional} bulk  dual ?}

As we have seen, the coordinates in \cite{Gubser:2016guj} reviewed above involve two parameters -- the radial coordinate $p^{n}$ and a p-adic number $x^{(n)}$. It is thus often regarded as a two dimensional space, while its boundary (the p-adic number) is regarded as a one dimensional $Q_p$ line. Here, we would like to arrange the coordinates so that the BT tree looks more analogous to AdS$_3$, and its boundary, to the complex plane $\mathbb{C}$.

\begin{figure}[htbp!]
\centering

\includegraphics[width=0.4\textwidth]{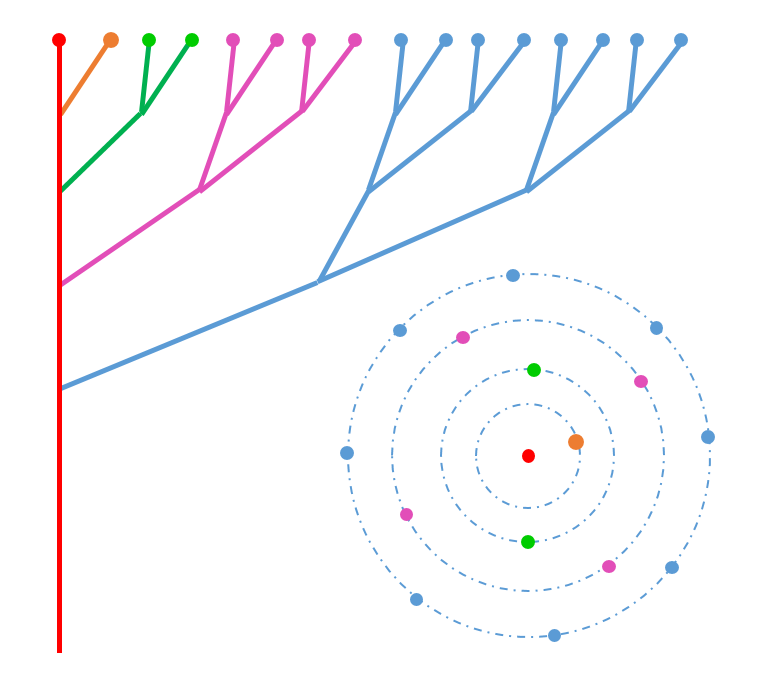}

\caption{The boundary of the BT tree can be viewed as a $2d$ plane. The boundary points of a same color brunch have the same norm and are on the same circle.}
\label{padicplane}
\end{figure}

A p-adic number is written as
\begin{eqnarray}
  x &=& p^v\sum_{i=0}^\infty a_i p^i,  \label{eq:pnumber}
\end{eqnarray}
with $a_i\in \{0,\dots,p-1\}$ and $a_0\neq 0$. And we have
\begin{eqnarray}
  |x|_p &=& |p^{v}|_p,\\
  |\sum_{i=0}^\infty a_i p^i|_p&=&1.
\end{eqnarray}
If we consider the complex plane, the holomorphic complex coordinate $\zeta$ can be written as
\begin{eqnarray}
  \zeta &\equiv& \rho e^{i \theta},
\end{eqnarray}
with
\begin{eqnarray}
  |\zeta|_R &=& |\rho|_R,\\
  |e^{i \theta}|_R &=& 1.
\end{eqnarray}
So $p^v$ plays the role of $\rho$, while $\sum_{i=0}^\infty a_i p^i$ plays the role of $e^{i \theta}$. Then the boundary of the BT tree can be regarded as a $2d$ plane with $0$ sitting on the origin, $|x|_p$ characterizing the distance from $x$ to origin, and $x$ with the same norm sitting on the same circle centered on the origin as shown in Fig. \ref{padicplane}. Since the boundary is a $2d$ plane, the BT tree itself is a $3d$ object.

The p-adic Poincare coordinates  can thus be taken as 
\be  \label{poincarecoord}
(z, \rho,e^{i\theta})  \equiv (p^{n+1}, p^v,  \sum_{i=0}^{n-v} a_i p^i),
\ee
where we have artificially taken out the unit norm piece in $x^{(n+1)}$ and take it as a third coordinate. 
This now looks more similar to the AdS$_3$ picture.

\subsection{BTZ black hole and the Black Hole coordinates}

In AdS$_3$, it is well known that the BTZ black hole can be understood as an orbifold of AdS by an appropriate Schottky group \cite{Maldacena:1998bw, Krasnov:2000zq}, since higher genus Riemann surfaces can be constructed as quotients of the projective complex plane $P(\mathbb{C})$. 
One can replace the complex numbers $\mathbb{C}$ by the p-adic numbers $Q_p$. By orbifolding the Bruhat-Tits tree by a  PGL$(2,Q_p)$ subgroup $\mathcal{S}$ with $n$ generators, one generates a genus $n$ graph. The $n=1$ case would be the closest cousin of the BTZ black hole. 
This was considered in \cite{Heydeman:2016ldy,Ebert:2019src}, and also in \cite{Hung:2019zsk} where we obtained the partition function of a p-adic CFT by covering the genus one graph with the tensor network that we have proposed.

\begin{figure}[htbp!]
\centering

\includegraphics[width=0.6\textwidth]{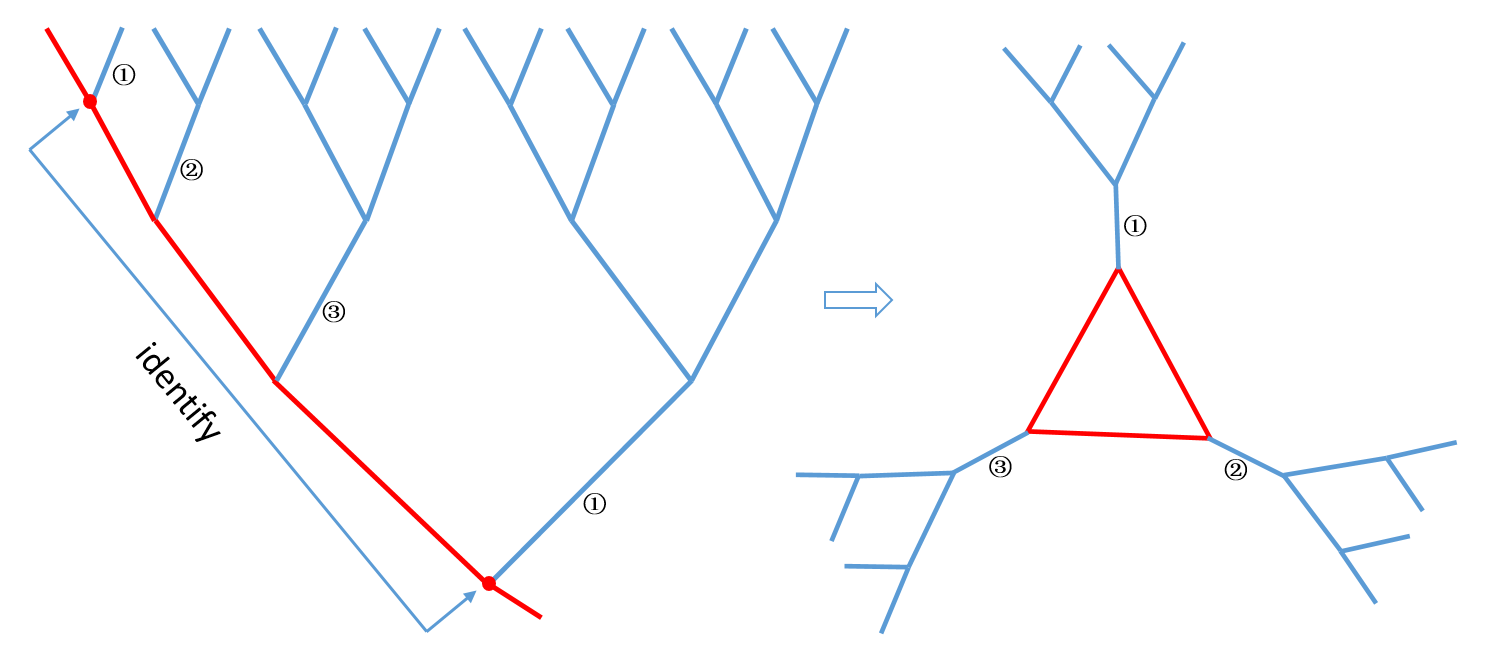}
\caption{Orbifold of the BT tree (at $p=2$) that converts it into a genus 1 graph. Here the periodicity $t$ =3.}
\label{fig:orbifold}
\end{figure}

In the case of a genus one graph, it is generated by quotienting the tree by a singly generated subgoup of PGL$(2, Q_p)$.
This subgroup actually has a very simple action on the tree. Consider picking a branch of the tree (such as the main-branch as shown in Fig. \ref{fig:orbifold}. ). The BT tree has translation invariance along this branch (when there is no cutoff surface). Each singly generated subgroup corresponds to translation on the chosen branch. The quotient is carried out by identifying every $t$ steps, and $t$ would play the role of a moduli of the resultant genus 1 curve. 
The resultant graph contains a loop and it is shown in Fig. \ref{fig:orbifold}.  
That closed loop has the same flavour as a horizon in the Euclidean BTZ black hole. As we are going to see the parallels run far deeper.  

An important point is that as soon as we consider any practical calculations, it is necessary to introduce a cutoff surface. The natural cutoff surface in this case appears to be one that is parallel to the horizon. This cutoff surface is thus very different from what is customarily used in say \cite{Gubser:2016guj}. We will call the coordinates there the Poincare coordinates and those cutoff surface a constant $z$ cutoff surface. 

Here we would like to assign a different set of coordinates so that the horizon and lines parallel to the horizon have constant radial coordinate. 

\begin{figure}[htbp!]
\centering
\subfigure[]{
\label{padicBH}
\includegraphics[width=0.35\textwidth]{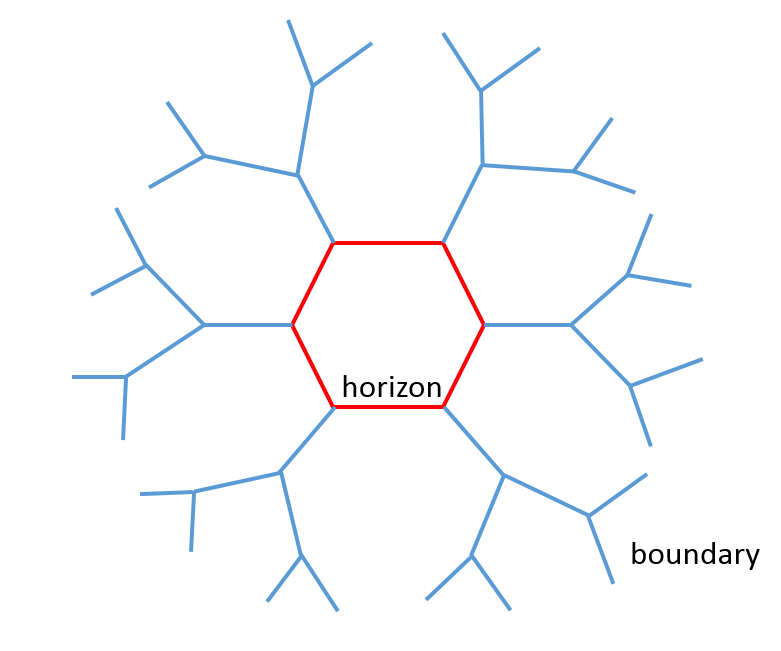}}
\subfigure[]{
\label{stripBH}
\includegraphics[width=0.6\textwidth]{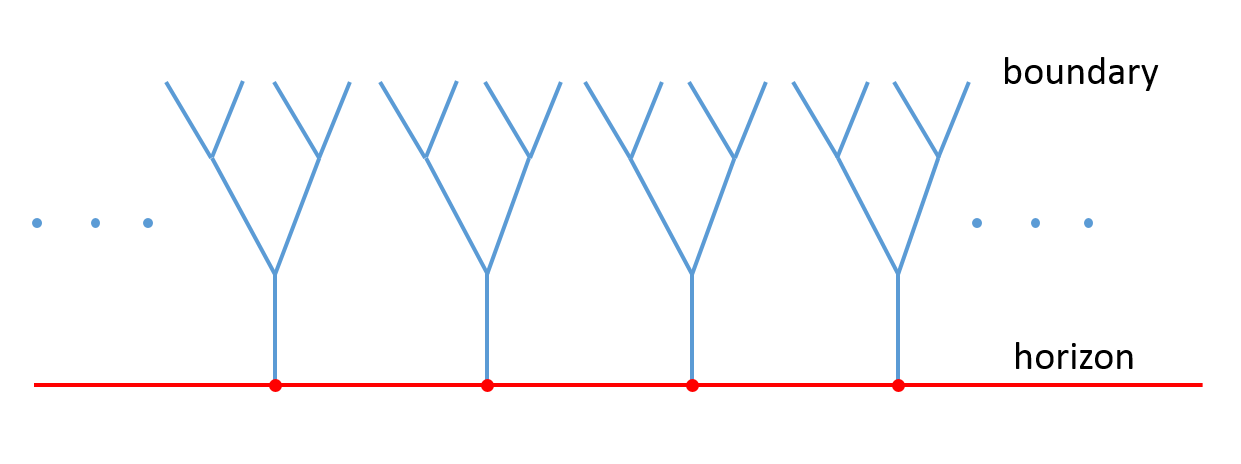}}

\caption{(a): The p-adic BTZ BH, (b): The cylinder type p-adic BH. The red curve is the horizon.}

\end{figure}

To proceed, it is convenient to consider the case where the periodicity of the loop approaches infinity. 
Without the cutoff surface, this would be the original BT tree, but now we are motivated to consider coordinates whose constant radial ``surfaces'' are parallel to the horizon. 
This uncompactified tree is shown in Fig. \ref{stripBH}.  

In the case of the usual BTZ black hole, when we uncompactify the spatial circle, the asymptotic boundary is a cylinder. Here, we will also refer to the uncompactified black hole as cylinder type. 

\begin{figure}[htbp!]
\centering
\includegraphics[width=0.8\textwidth]{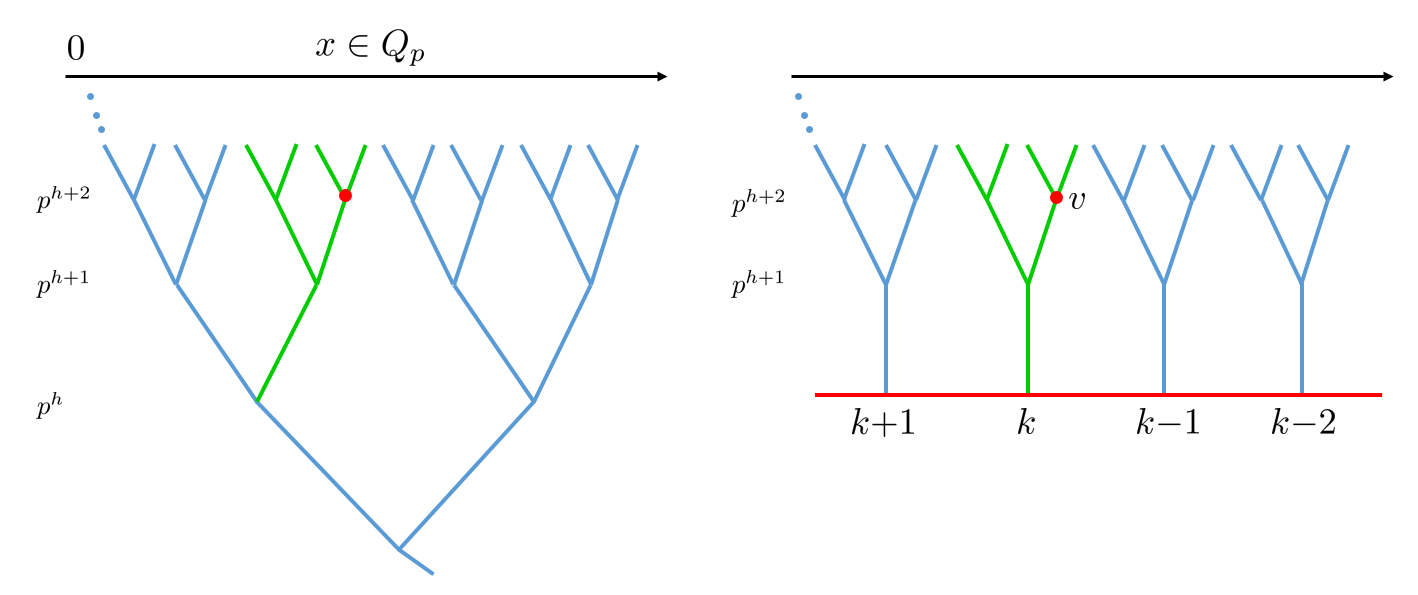}

\caption{The coordinates on the cylinder type BH. Each branch of the BH is taken from the BT tree emanating from the vertex labeled $(p^h,0)$ in Poincare coordinates, which is colored green in the figure. The coordinates of $v$ is $(k,p^{h+2},y^{(h+2)})$.}
\label{BHcoord}
\end{figure}

The cylinder type BH has translation invariance along its horizon. At every vertex along the horizon an identical subtree emanates from the vertex.  Our black hole coordinates are chosen as follows. We label every vertex on the horizon by an integer $k$. 
Supposing each branch emanating from each vertex is a semi-infinite tree if not for the cut-off surface.  
Here, consider that it is cut from the BT tree at $(p^h,0)$. We can then label the vertices on one branch by borrowing from the Poincare coordinates as
\begin{eqnarray}
  (p^{h+l},y^{(h+l)}) &=& (p^{h+l},p^h\sum_{i=0}^{l-1}a_i p^i). \;\;\;\;\;(a_0\neq 0,\;l\geq 1)
\end{eqnarray}
Then the coordinates on the cylinder type BH is given by $(k,p^{h+l},y^{(h+l)})$ as shown in Fig. \ref{BHcoord}.

i.e. the second and third entries came from Poincare labels originally assigned to that branch. This same branch is repeated an infinite number of times on the horizon and they are distinguished by the label $k$. 
See Fig. \ref{BHcoord}, where the branch colored green on the horizon can be thought of as being taken from the BT tree with Poincare labels. This gives a unique label to every vertex on the tree. 

One must have noticed that a parameter $h$ has been introduced. As we are going to show when we compute correlation functions, that $h$ controls the temperature of the p-adic CFT. We will give further support to this claim when we consider the Chern-Simons formulation of the tensor network, and show that $h$ would appear precisely as a moduli in the connection that plays the same role of parametrising the horizon radius. 

These labels do not look similar to the black hole coordinates on AdS$_3$ yet. To make them look more familiar,  we need to introduce a function that we call $\Theta$. It is essentially the analogue of logarithm that takes a p-adic number of unit norm to another p-adic number that is however periodically identified.

\subsubsection{An alternative p-adic ``logarithm'' and ``exponential''}

Before proceeding to give the precise definition of the black hole coordinates, we were inspired by the BT tree itself, which inspired us to consider an alternative notion of  ``logorithm''  for p-adic numbers that we denote as  $\Theta$. From which it is also going to inspire an alternative notion of ``exponential''. 

{\bf \underline{An alternative ``logarithm''}}

For the usual logarithm that acts on complex numbers, $\Theta$ takes a pure phase $y$ with norm $|y|_R=1$ to a number $\theta$ that is periodically identified every $2\pi$ (alternatively we call that a branch-cut):
\begin{eqnarray} 
\Theta: y=e^{i\theta}\rightarrow \theta, \label{eq:cond1}
\end{eqnarray}
satisfying
\begin{eqnarray}
  \Theta(y_1)-\Theta(y_2) &=& \Theta\left(y_1/y_2\right), \label{eq:cond2}
\end{eqnarray}
since
\begin{eqnarray}
e^{i(\theta_1-\theta_2)} &=& \frac{e^{i\theta_1}}{e^{i\theta_2}}. \label{eq:cond1}
\end{eqnarray}

Now analogously,  let's define the p-adic version of the logarithm $\Theta$ as a map from a p-adic number $y$ with norm $|y|_p=1$ to a periodic number $\theta$:
\begin{eqnarray}
\Theta: y=\sum_{i=0}^\infty a_i p^i \rightarrow \theta,\;\;\;(a_0\neq 0)
\end{eqnarray}
satisfying
\begin{eqnarray}
\label{theta}
  \Theta(y_1)-\Theta(y_2) &=& \Theta\left(y_1/y_2\right).
\end{eqnarray}

Here we consider a simple example to illustrate what $\Theta$ does. Let's choose $p=5$ and truncate $y$ so that we keep only the leading term $y=a_0 p^0 + \mathcal{O}(p)$. Then  $y \in \{1,2,3,4\}$. We find that
\begin{eqnarray}
  1,2,3,4 &\xrightarrow{\div 1}& 1,2,3,4,\\
  1,2,3,4 &\xrightarrow{\div 2}& 3,1,4,2,\\
  1,2,3,4 &\xrightarrow{\div 3}& 2,4,1,3,\\
  1,2,3,4 &\xrightarrow{\div 4}& 4,3,2,1.
\end{eqnarray}
Note that the division computed above are p-adic divisions where the result has to be arranged in terms of the expansion (\ref{eq:pnumber}) and then from which one extracts $a_0$.

\begin{figure}[htbp!]
\centering

\includegraphics[width=0.3\textwidth]{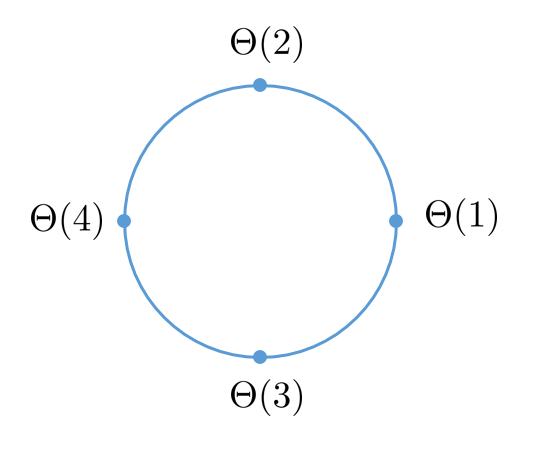}

\caption{The $\Theta(i)$ on the circle.}
\label{thetacircle}
\end{figure}

To satisfy (\ref{theta}), we can choose
\begin{eqnarray}
  \Theta(1)=0,\;\;\;\;\Theta(2)=\frac{\pi}{2},\;\;\;\;
  \Theta(3)=\frac{3\pi}{2},\;\;\;\;\Theta(4)=\pi,
\end{eqnarray}
with the period $2\pi$ as shown in Fig. \ref{thetacircle}.

For a general $p$ and any truncation of $y$, there always exists a map $\Theta$ satisfying (\ref{theta}), though the explicit expression of $\Theta$ hasn't been worked out. 

And we can further define a map $e_p$ as the inverse of $\Theta$, i.e.
\begin{eqnarray}
\label{ep}
  e_p(\Theta(y)) &=& y.
\end{eqnarray}

This function is almost an exponential. Now we would like to introduce a notion of the exponential that acts on an arbitrary p-adic number.

{\bf \underline {An alternative  ``exponential''}}

Now we can define the p-adic exponential map $\mathcal{E}_p$ as
\begin{eqnarray}
  \mathcal{E}_p:  w=(\phi,\tau)&\rightarrow& p^\phi e_p(\tau).  \label{eq:exponential}
\end{eqnarray}
where $\phi$ is an integer, and $\tau$ a periodic number in the image of $\Theta$.
 The calculation rules of $w$ are simply
\begin{eqnarray}
  (\phi_0,\tau_0)-(\phi_1,\tau_1) &=& (\phi_0-\phi_1,\tau_0-\tau_1),\\
  a(\phi,\tau)&=&(a\phi,a\tau).
\end{eqnarray}
The map $\mathcal{E}_p$ has following properties:
\begin{eqnarray}
\mathcal{E}_p((0,0))&=&1,     \label{eq:exp1}\\ 
  \mathcal{E}_p(w_0)\mathcal{E}_p(w_1) &=& p^{\phi_0}p^{\phi_1}e_p(\Theta(y_0))e_p(\Theta(y_1))   \\
  &=&p^{\phi_0+\phi_1}e_p(\Theta(y_0y_1))\\
  &=&p^{\phi_0+\phi_1}e_p(\Theta(y_0)+\Theta(y_1))\\
  &=& \mathcal{E}_p(w_0+w_1),   \label{eq:exp2}
\end{eqnarray}
where we have used (\ref{theta}),(\ref{ep}).

We further define $\sinh_p,\cosh_p$ as
\begin{eqnarray}
\label{sinh}
  \sinh_p(w) &\equiv& \frac{\mathcal{E}_p(w)-\mathcal{E}_p(-w)}{2},\\
  \label{cosh}
  \cosh_p(w) &\equiv& \frac{\mathcal{E}_p(w)+\mathcal{E}_p(-w)}{2}.
\end{eqnarray}
And we have
\begin{eqnarray}
\label{identity}
  \cosh^2_p(w)-\sinh^2_p(w) &=& 1. 
\end{eqnarray}

Staring at the requirements (\ref{eq:cond1}, \ref{eq:cond2}, \ref{eq:exp1}, \ref{eq:exp2}) one would have been reminded of solutions of characters of representations in the principal series of PGL$(2,Q_p)$. We note that one very important difference is that the current map takes p-adic numbers to an ordered series where we have borrowed labels from the real line $2\pi/k$ to rename the results and to act as a mnemonic of the number of steps i.e. $\Theta(y_1)- \Theta(y_2)$ between the images in the ordered series. This is vastly different from the usual exponential whose action is defined based on its power series expansion. Note in particular that there is no divergence here that otherwise plagued the usual exponential (or logarithm for that matter).
As we are going to see, these definitions are inspired by the computation of correlation functions that first follows from the BT tree with these different choices of cut-off surface. 

\subsubsection{A black hole coordinates}

\begin{figure}[htbp!]
\centering

\includegraphics[width=0.75\textwidth]{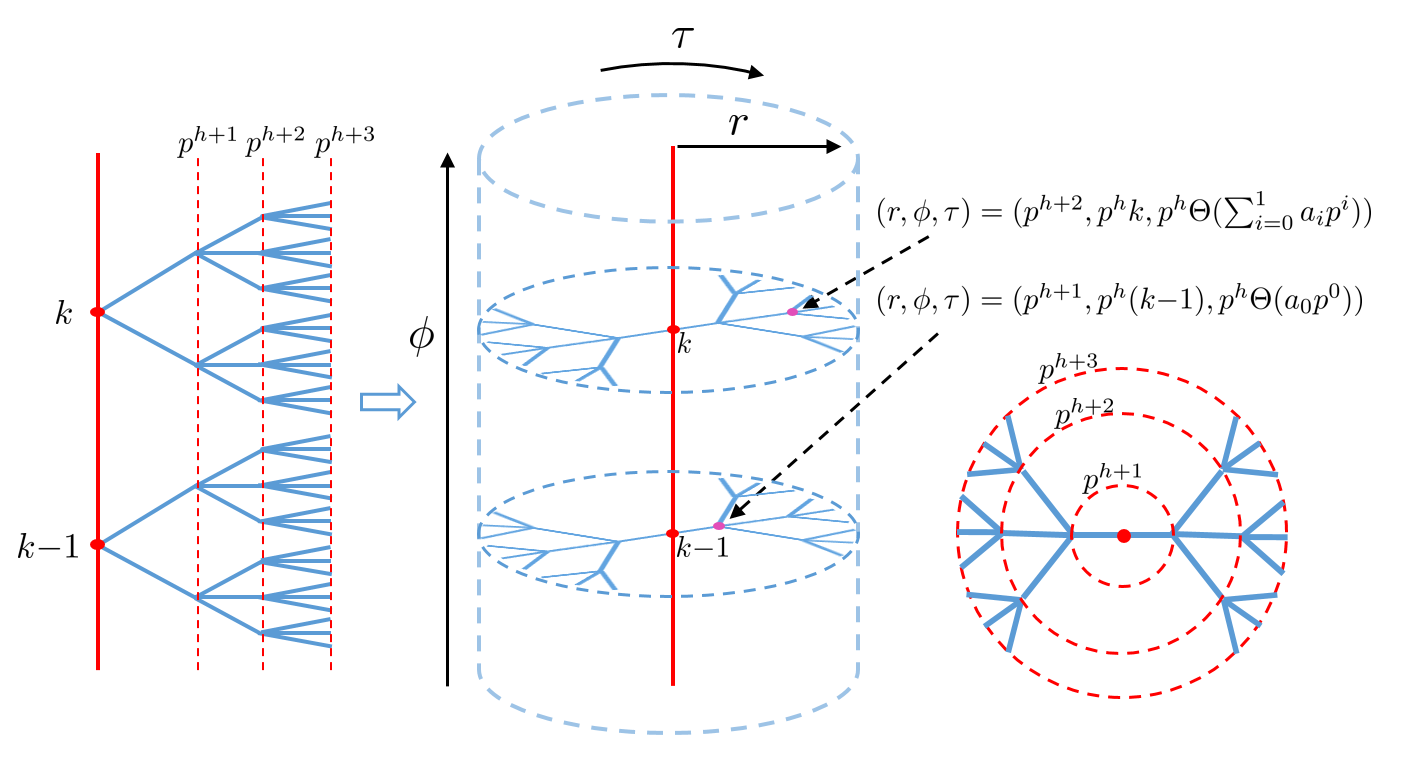}

\caption{The cylinder type BH looks like a cylinder ($p=3$). The red line is its horizon. The right most picture shows the cross-section of the cylinder.}
\label{cylinder}
\end{figure}

In the previous sub-section, we argued that the p-adic tree can be naturally treated as a 3d object that mimics AdS$_3$ closely. 
In the case of the cylinder type BH one can also arrange the Bruhat-Tits tree into a cylinder like object as shown in Fig. \ref{cylinder}. 
Recall that we introduced an alternative coordinates to label vertices on the cylinder.
\begin{eqnarray}  
  (k,p^{h+l},y^{(h+l)}) &=& (k,p^{h+l},p^h\sum_{i=0}^{l-1}a_i p^i). \;\;\;\;\;(a_0\neq 0,\;l\geq 1)  \label{eq:bhcoord1}
\end{eqnarray}

The black hole coordinates we would like to introduce that makes use of the {\it logarithm} introduced in the previous sub-section is related to the above as follows
\begin{eqnarray}
\label{globalcoord}
  (r,\phi,\tau) &=&(p^{h+l},p^h k,p^h\Theta
  (\sum^{l-1}_{i=0}a_ip^i)).  \label{eq:rphitau}
\end{eqnarray}
Here $r$ is the radial coordinate, and $\phi,\tau$ are coordinates on the cylinder with $\tau$ the compactified direction as shown in Fig. \ref{cylinder}.

\subsection{Local diffeomorphism connecting black hole and Poincare coordinates}

The coordinate transformation between black hole coordinate (\ref{globalcoord}) and Poincare coordinate (\ref{poincarecoord}) is simply given by
\begin{eqnarray}
  \theta &=& \tau/p^h,\\
  \rho&=&p^{\phi/p^h},  \label{eq:norm}\\
  \label{xrule}
  x&\equiv& \rho e_p(\theta) =p^{\phi/p^h} e_p(\tau/p^h) ,\\
  \label{zandr}
  z&=&r p^{\phi/p^h}/p^h.
\end{eqnarray}
Here we have used $l-1=n-v$, and chosen $k=v$. Note that $\theta$ is indeed always in the image of the function $\Theta$ from the definition (\ref{globalcoord}).
Let's introduce $w=(\phi,\tau)$, and the p-adic black hole coordinate can be written as $(r,w)$.  Using the definition of the ``exponential'' defined in (\ref{eq:exponential}), we can write
\be
x = \mathcal{E}_p(\frac{w}{p^h}).
\ee

\begin{figure}[htbp!]
\centering

\includegraphics[width=0.75\textwidth]{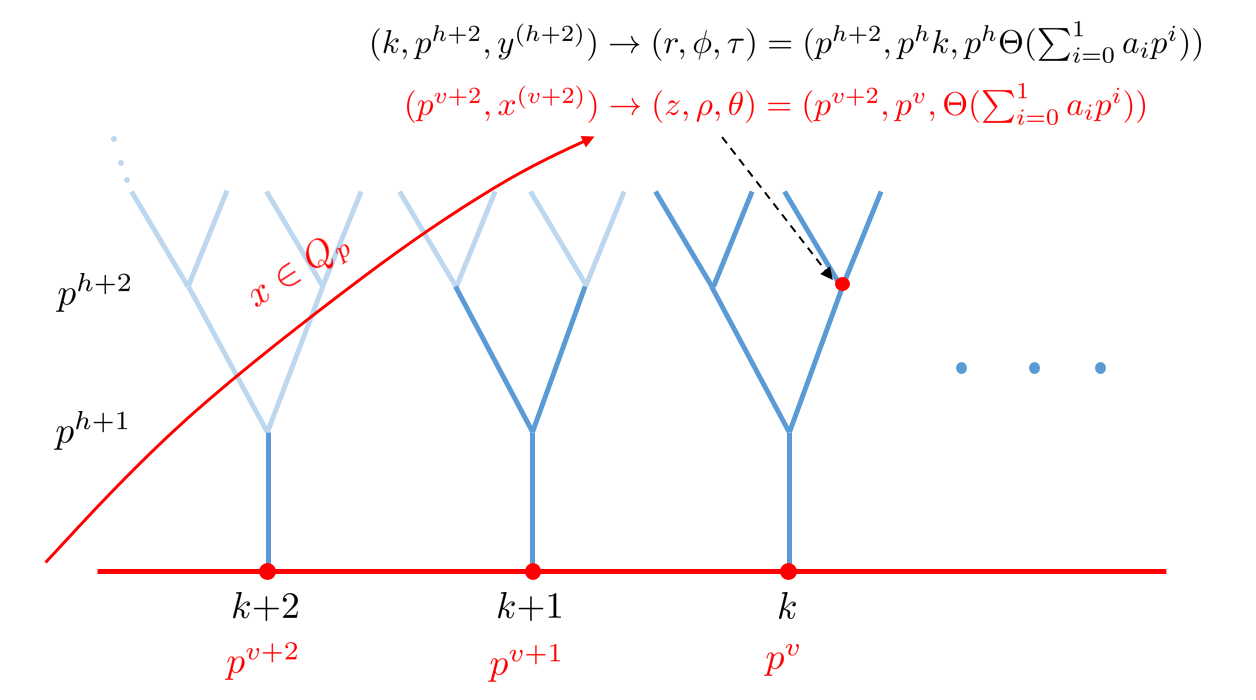}

\caption{A marked vertex in Black Hole (in black) and Poincare (in red) coordinates. The red arrow gives the constant Poincare radial coordinate $z$ surface. The horizon here is the "main branch" in Poincare coordinates and on which marks the radial $z$ values.   }
\label{fig:compare0}
\end{figure}

To illustrate the coordinate transformations more clearly, see Fig. \ref{fig:compare0}.

This coordinate transformation can be contrasted with the transformation between Poincare coordinates and the BTZ coordinates \cite{Maldacena:1998bw}. 

There, we have
\begin{eqnarray}
\zeta &=& \sqrt{\frac{r^2-r_h^2}{r^2} }  \exp{(\frac{r_h}{l} \phi+ i \tau)}\\
z & =& \frac{r}{r_h}\exp{ (\frac{r_h}{l} \phi)},
\end{eqnarray}
where here we have already set the angular momentum of the BTZ black hole to zero keeping only the outer-horizon, which is the situation analogous to the cylinder type Bruhat Tits tree black hole. The  metrics corresponding to these coordinates take the form
\begin{eqnarray}
ds_{Poincare}^2 &=& l^2 \frac{dz^2}{z^2}+ z^2 d\omega d\bar\omega, \\
ds_{BH}^2 &=& \frac{(r^2- r_h^2)}{r^2} d\tau^2 + \frac{r^2 l^2}{r^2-r_h^2} dr^2 + r^2 d\phi^2.
\end{eqnarray}

Apart from the choice of norms of $\zeta$ being somewhat different from (\ref{eq:norm}), the relations are step by step identical.
Moreover, this distinction in the choice of normalization along the "CFT" directions disappears towards the asymptotic boundary $r\to\infty$, where the extra factor of $r$ reduces to unity. This means that the interpretation as far as the CFT is concerned would be identical. 
In the following we will show that, indeed, the form of the correlation functions following from the bulk computations with the black hole cut-off surface can be interpreted as a finite temperature correlator in analogy to the usual CFT on a cylinder.

\subsection{Correlation functions in the BTZ background }
Having assigned a new black hole coordinate to the Bruhat Tits tree and pick a corresponding cutoff surface parallel to the purported horizon, we can now compute boundary CFT correlation functions using the bulk. 
Here, we will compute the CFT correlation functions using the tensor network we have proposed in \cite{Hung:2019zsk}. 
A very brief review of p-adic CFT and our tensor network can be found also in the first of this series \cite{paper1, paper2}. 

{\bf Faster than lightening review of the tensor network}

Here, all we will need in the following is that we are covering the BT tree with a tensor network, with every vertex hosting a tensor and each edge emanating from a vertex is an index of the vertex tensor. 
Being a $p+1$-valent tree, each tensor also has $p+1$ indices. It is explicitly given by
\be
 T^{a_1 a_2 \cdots a_{p+1}} = \sum_{b_1\cdots b_{p-2}} C^{a_1 a_2 b_1} C^{b_1 a_3 b_2}  \cdots C^{b_{p-2} a_p a_{p+1}},
\ee
where $C^{abc}$ are OPE coefficients of the boundary CFT, with a collection of primary fields labeled by $\{a,b,c \cdots\}$ and corresponding conformal dimensions $\{\Delta_a, \Delta_b, \Delta_c \cdots\}$.
Two vertices connected by an edge denote a contracted index between two vertex tensors. The contraction is a weighted sum, with each term with label $a$ weighted by $p^{-\Delta_a}$. 
The CFT partition function is defined as the evaluation of the tensor network covering the BT tree, such that the un-contracted dangling legs at the cutoff surface is projected to the identity label $1$, the unique identity operator with zero conformal dimension.

Each operator $\mathcal{O}_a$ inserted at point $x$ would then correspond to projecting the leg $x$ to label $a$. The evaluation of the tensor network with appropriate boundary conditions thus returns for us the correlation functions of the CFT. 

One can also define bulk operator $\phi_a$ inserted at a bulk vertex $v$. This is achieved by fusing a label $a$ to the fusion tree at $v$ defined by the OPE coefficients $C^{abc}$. For further details one can refer to volume one of our current series \cite{paper2}. 

{\bf Correlation functions}

Now we are ready to compute correlation functions. Let us remark that while the result here is based on the tensor network, the form of correlation functions up to three point functions are completely universal. 

\begin{figure}[htbp!]
\centering

\includegraphics[width=0.6\textwidth]{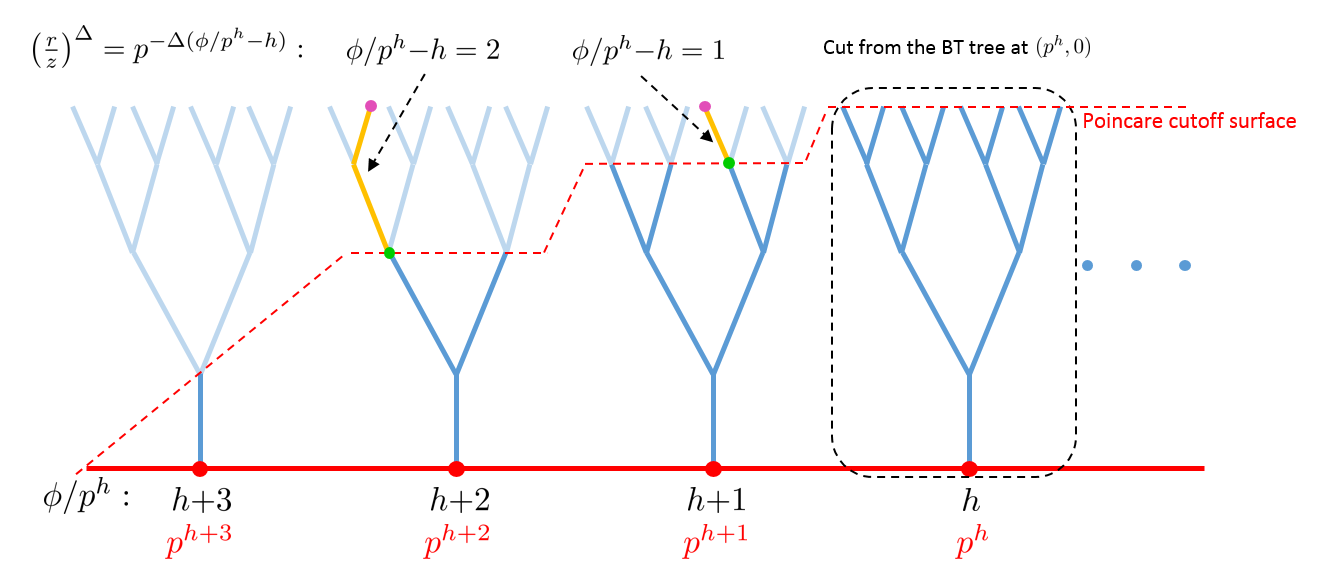}

\caption{Computing 2 point functions in the cylinder type black hole background. }
\label{fig:2pt}
\end{figure}

Consider two points $v_0=(r_0,\phi_0,\tau_0)$ and $v_1=(r_1,\phi_1,\tau_1)$ in black hole coordinates. Their two point function is given by:
\be
\label{phiphi}
  \langle\phi^a(v_0)\phi^a(v_1)\rangle = p^{-\Delta_a d(v_0,v_1)}, 
\ee
with $d(v_0,v_1)$ the distance between $v_0,v_1$.
If we would like to compute correlation of the p-adic CFT, we should consider putting these operators on the cutoff surface at constant $r_0$, and then finally pushing $r_0$ to infinity. 
\be
  \label{OO}
  \langle O^a(\phi_0,\tau_0)O^a(\phi_1,\tau_1)\rangle\equiv\lim_{r_0\to\infty}
  r_0^{2 \Delta_a} \langle\phi^a(v_0)\phi^a(v_1)\rangle,
\ee
where
 $(\phi_0,\tau_0),(\phi_1,\tau_1)$ are the boundary points in black hole coordinate. 
This is illustrated in Fig. \ref{fig:2pt} when $r_1 = r_0$. The factors of $r_0$ are normalizations that take away the divergence in the geodesic distance as one pushes the cutoff surface to infinity. This is analogous to the usual practice in AdS/CFT, and has been discussed in detail before in \cite{Hung:2019zsk}.

When the dust settles, we find
\be
\langle O^a(w_0)O^a(w_1)\rangle =\frac{1}
{|2p^h\sinh_p(\frac{w_0-w_1}{2p^h})|_p^{2\Delta_a}}.  \label{eq:2ptbh}
\ee

The result now appears uncannily similar to the correlation functions of real CFTs on a cylinder. In fact it is in the process of understanding this result that had inspired the definition of the ``exponential''. 

It may appear somewhat mysterious how (\ref{OO}) reduces to (\ref{eq:2ptbh}). Let us make it explicit by inspecting its relationship with correlation functions computed in Poincare coordinates (where the asymptotic boundary is ``plane like'' as opposed to ``cylinder like'').

\subsection{Weyl transformation in the p-adic CFT}

Recall that in Poincare coordinate, we have $\tilde{v}_0=(z_0,x_0)$ and $\tilde{v}_1=(z_1,x_1)$.
Recall that in Poincare coordinate, we have \cite{Gubser:2016guj}
\begin{eqnarray}
  \langle\phi^a(\tilde{v}_0)\phi^a(\tilde{v}_1)\rangle&=&p^{-\Delta_a d(\tilde{v}_0,\tilde{v}_1)}=
  \frac{z_0^{-\Delta_a}z_1^{-\Delta_a}}
  {|x_0-x_1|_p^{2\Delta_a}},\\
  \langle \tilde{O}^a(x_0)\tilde{O}^a(x_1)\rangle &\equiv& \lim_{z_i\to\infty}z_0^{\Delta_a}z_1^{\Delta_a}\langle\phi^a(\tilde{v}_0)\phi^a(\tilde{v}_1)\rangle
  =\frac{1}
  {|x_0-x_1|_p^{2\Delta_a}},
\end{eqnarray}
with $x_0, x_1$ the boundary points in Poincare coordinate. We have denoted operators inserted along the constant radial cutoff surface in {\it Poincare coordinates} by $\tilde{\mathcal{O}}_a$, with an extra tilde, to distinguish them from operators inserted at constant $r$ surfaces in the black hole coordinates. 
It is evident that $d(v_0,v_1)=d(\tilde{v}_0,\tilde{v}_1)$. 
Now when we want to compute correlation functions in the boundary "cylinder frame",  we have to insert operators at the cutoff surface $r_0$ as opposed to constant $z$ surface. Therefore we need to ``rectify" the difference in the geodesic distance when we push the operators at the constant $z$ surface to the constant $r_0$ surface.  This is illustrated in Fig \ref{fig:2pt}. 
A moment's thought suggests that the rectification corresponds to change in the normalizations, and we have
\begin{eqnarray}
\label{OO3}
\langle O^a(\phi_0,\tau_0)O^a(\phi_1,\tau_1)\rangle &=& r_0^{\Delta_a}r_1^{\Delta_a}z_0^{-\Delta_a}z_1^{-\Delta_a}
\langle \tilde{O}^a(x_0)\tilde{O}^a(x_1)\rangle=
\frac{p^{2h\Delta_a}|x_0|^{\Delta_a}_p|x_1|^{\Delta_a}_p}
{|x_0-x_1|_p^{2\Delta_a}},
\end{eqnarray}
where we have used (\ref{xrule}),(\ref{zandr}).
Using the $w$ coordinate and $x=\mathcal{E}_p(w/p^h)$,
(\ref{OO3}) becomes
\begin{eqnarray}
\nonumber
\langle O^a(w_0)O^a(w_1)\rangle &=&\frac{|p^{-h}|_p^{2\Delta_a}|\mathcal{E}_p(w_0/p^h)|_p^{\Delta_a}|\mathcal{E}_p(w_1/p^h)|_p^{\Delta_a}}
{|\mathcal{E}_p(w_0/p^h)-\mathcal{E}_p(w_1/p^h)|_p^{2\Delta_a}}\\
\label{OO4}
&=&\frac{1}
{|2p^h\sinh_p(\frac{w_0-w_1}{2p^h})|_p^{2\Delta_a}}.
\end{eqnarray}

Using (\ref{OO4}),(\ref{OO}), we note also that the bulk two point function (\ref{phiphi}) can also be expressed as
\begin{eqnarray}
\nonumber
  \langle\phi^a(v_1)\phi^a(v_2)\rangle &=&\frac{r_1^{-\Delta_a} r_2^{-\Delta_a}}
{|2p^h\sinh_p(\frac{w_1-w_2}{2p^h})|_p^{2\Delta_a}}\\
\label{phiphi2}
&=&\frac{p^{-(2h+l_1+l_2)\Delta_a}} 
{|2p^h\sinh_p(\frac{w_1-w_2}{2p^h})|_p^{2\Delta_a}}.
\end{eqnarray}

These results are very suggestive that the black hole frame is describing a finite temperature boundary p-adic CFT. 
This can be made most transparent when we compare with correlation functions of real 2d CFT at temperature $T\equiv 2\pi/\beta$. This is equivalent to evaluating correlation functions on a cylinder coordinate. We suggestively take the holomorphic complex coordinate on the cylinder also as $w$. The planar holomorphic complex coordinate $\zeta$ is related to the cylinder coordinate $w$ by $\zeta=e^{2\pi w/\beta}$. 
Correlation functions on the cylinder are then related to correlation functions on the plane by
\begin{eqnarray}
\nonumber
  \langle O(w_0)O(w_1)\rangle &=&
  |z_0^\prime|^\Delta
  |z_1^\prime|^\Delta
  \langle \tilde{O}(z_0)\tilde{O}(z_1)\rangle
  =
  \frac{|(e^{2\pi w_0/\beta})^\prime|^\Delta
  |(e^{2\pi w_1/\beta})^\prime|^\Delta}
  {|e^{2\pi w_0/\beta}-e^{2\pi w_1/\beta}|^{2\Delta}}\\
 \nonumber
  &=&\frac{|\frac{2\pi}{\beta}|^{2\Delta}|e^{2\pi w_0/\beta}|^\Delta
  |e^{2\pi w_1/\beta}|^\Delta}{|e^{2\pi w_0/\beta}-e^{2\pi w_1/\beta}|^{2\Delta}}\\
  \label{OOCFT1}
  &=&\frac{1}{\left|\frac{\beta}{\pi}\sinh\left(\frac{\pi(w_0-w_1)}{\beta}\right)\right|^{2\Delta}}.
\end{eqnarray}
with $|\dots|$ the real norm $|\dots|_R$. Comparing (\ref{OO4}) and (\ref{OOCFT1}), we find
\begin{eqnarray}
  \exp(\dots) &\rightarrow& \mathcal{E}_p(\dots),\\
  \label{betah}
   \frac{2\pi}{\beta} &\rightarrow& \mathcal{T}  = p^{-h},  \label{eq:Tph} \\
   |\dots|_R&\rightarrow&|\dots|_p.
\end{eqnarray}
Note that we denote the p-adic temperature which is a p-adic number, by $\mathcal{T}$.

Now we have more confidence to regard $\mathcal{E}_p$ as the appropriate exponential function here. In $\textrm{CFT}_2$, $\zeta=e^{2\pi w/\beta}$ maps a CFT with temperature $|2\pi/\beta|_R$  to a zero temperature CFT, while $x=\mathcal{E}_p(w/p^h)$ maps a p-adic CFT with temperature $|p^{-h}|_p$ to a zero temperature p-adic CFT. The extra parameter that is introduced along with the black hole coordinates in (\ref{globalcoord}) acquires a meaning of a temperature! 
We will comment further about black hole thermodynamics after our discussion of the Wilson lines.

Now we are ready to connect correlation functions on the ``planar frame'' with correlation functions on the ``finite temperature cylinder frame'' more generally. As discussed above, they are related by the ``correction" that one has to make because of the change in the cutoff surface. Therefore, each operator insertion would be corrected by the same set of factors $ |z_i^\prime|^\Delta$ as discussed above. We can thus summarise the general rule of transformations of correlation functions under general Weyl transformation under coordinate transformation $x \to f_p(x)$:
\begin{eqnarray}
  O(x) &=& |f_p^J(x)|_p^\Delta \tilde{O}(f(x)),
\end{eqnarray}
where $f_p^J$ is fully determined by $f_p$, with the caveat that its general expression for arbitrary cutoff is yet to be found. 
Similar to $e(w)\equiv e^{w}$ satisfying $e^J(aw)=a\;e(aw)$, our $\mathcal{E}_p$ satisfies
\begin{eqnarray}
  \mathcal{E}_p^J(a w) &=& a\;\mathcal{E}_p(a w),
\end{eqnarray}
which can be obtained by observing (\ref{OO4}).

This is completely parallel to the story in usual two dimensional CFTs, where under Weyl transformation $f(w)$, a primary operator $O$ with conformal dimension $\Delta$ transforms as
\begin{eqnarray}
  O(w) &=& |f^J(w)|^\Delta \tilde{O}(f(w)),
\end{eqnarray}
with $f^J(w)\equiv\partial f/\partial w$.

\section{The Chern-Simons formulation and the BTZ connection on the Bruhat-Tits tree}

In \cite{Hung:2018mcn}, we showed that it is possible to formulate a Chern-Simons like theory on the Bruhat-Tits tree. While
it is not immediately obvious what is the appropriate action for a topological gauge theory on a tree graph, it is quite apparent
that if it existed, one would also impose gauge invariance at every vertex where Wilson lines meet. While the tree graph has no loops and therefore it is not clear how to define generic ``flat connections'', a pure gauge connection that can be written as $g(v_1) g^{-1}(v_2)$ for some gauge group valued scalar function $g(v)$
would certainly qualify for being flat. 

Taking these two assumptions, we constructed a Wilson line network of a purported PGL$(2,Q_p)$ gauge group with a connection that can be written as a pure gauge \cite{Hung:2018mcn}. 
The form of the flat connection ended up looking completely parallel to the SL$(2,\mathbb{C})$  flat connection that follows from the Euclidean AdS$_3$ metric \footnote{In the Lorentzian case the gauge group is SL$(2, \mathbb{R}) \times $ SL$(2,\mathbb{R})$.}. Moreover the Wilson line junctions evaluated in this flat connection coincides with the tensor network that we constructed to recover the p-adic partition function. 

It is well known that 2d CFT states with finite energy density and  momenta are dual to BTZ black holes whose metrics can be expressed as flat connections of SL$(2,\mathbb{C})$ in the Chern-Simons formulation. 
It would immediately suggest that the BTZ connection perhaps can be transplanted to the p-adic Chern-Simons theory as well.

This is indeed possible -- which is the main subject of the section. In fact it is this computation that inspired us to look for a different set of coordinates so that the parallel between the p-adic AdS/CFT and AdS$_3$/CFT$_2$ is more complete.

\subsection{The pure AdS$_3$ connection vs Bruhat-Tits connection }

To introduce the p-adic BTZ connection it is helpful to start with the EAdS$_3$ BTZ black hole expressed as a flat connection in Chern-Simons gauge theory, a classic story instigated in \cite{Witten:1988hc}.
In this paper we will make use of the notations and expressions set up in  \cite{Carlip:2005zn}. 

For a generic asymptotically AdS$_3$ spacetime, the (Euclidean) metric can be expressed as
\begin{eqnarray}\label{metric}
  ds^2 &=& -4Gl(L(\zeta)d\zeta^2+\bar{L}(\bar{\zeta})d\bar{\zeta}^2) +l^2d r^2+
  (l^2e^{2 r}+16G^2L(\zeta)\bar{L}(\bar{\zeta})e^{-2r})d\zeta d\bar{\zeta}.
\end{eqnarray}
Here $\zeta$ is related to the cylinder coordinates $(\phi,\tau)$ by $\zeta = \phi + i \tau$.
The vierbein $e^a_\mu$ satisfying $g_{\mu\nu} = e^a_{\mu} e^b_{\nu}\delta_{ab}$ together with its derived spin connection $\omega^{ab}_\mu$ can be expressed as a SL$(2,\mathbb{C})$ Chern-Simons connection which takes the form \footnote{We have converted the $SL(2,\mathbb{R}) \times SL(2,\mathbb{R})$ connection into an SL$(2,\mathbb{C})$ connection. We presented the ``holomorphic'' part of the connection which is given as in (\ref{eq:Acon}). } 
 \begin{equation} \label{eq:Acon}
 A=
 \left( 
 \begin{array}{ccc} 
-\frac{1}{2}dr & e^{r}d\zeta \\ 
\frac{4G}{l}L(\zeta)e^{-r}d\zeta & \frac{1}{2}dr \\ 
 \end{array} \right). 
 \end{equation}
So we have
\begin{equation}
A_\zeta=
  \left( 
 \begin{array}{ccc} 
0 & e^{r} \\ 
\frac{4G}{l}L(\zeta)e^{-r} & 0 \\ 
 \end{array} \right),\;\; A_r =
 \left( 
 \begin{array}{ccc} 
-\frac{1}{2} & 0 \\ 
0 & \frac{1}{2} \\ 
 \end{array} \right).
\end{equation}

A finite Wilson line stretching from $v_1=(r_1,\zeta_1)$ to  $v_2=(r_2,\zeta_2)$  is given by
\begin{eqnarray}
  \mathfrak{W}(v_1\to v_2) &=&
  P\;\textrm{exp}\left(\int_{v_1}^{v_2}A_\mu(\xi) d\xi^\mu\right)\nonumber\\
  &=&P\;\textrm{exp}\left(\int_{\zeta_1}^{\zeta_2}A_\zeta (r_1,\zeta)d\zeta\right) \cdot P\;\textrm{exp}\left(\int_{r_1}^{r_2}A_r(r,\zeta_2)dr \right).  \label{eq:adsA}
\end{eqnarray}
Here we have chosen a specific path for convenience, but of course the result is path-independent since the connection is flat.

In the case where $L(\zeta) = 0$, the connection describes pure AdS space. It is known that this connection takes a simple form
\be
A(r,\zeta) = g^{-1} d g, \qquad g \in SL(2,\mathbb{C}), \qquad g(r,\zeta) = \left(\begin{tabular}{cc} $e^{-r}$ & $\zeta$ \\
0 & 1
\end{tabular}\right)
\ee

Then (\ref{eq:adsA}) evaluates to

\begin{equation}
\mathfrak{W}(v_1\to v_2)\vert_{L(\zeta)=0}=\left(
\begin{array}{cc}
 e^{-\delta r}  & e^{r_1 }\delta \zeta \\
 0 & 1 \\
\end{array}
\right)e^{\delta r/2}. \label{eq:wilads3}
\end{equation}
where 
$k=l/4G$, $\delta \zeta=\zeta_2-\zeta_1$, $\delta r=r_2-r_1$. And we set $k=1$ for convenience.

The ``pure'' BT tree connection was obtained through a series of guess. But when the dust settles, it can be very simply summarised  most transparently as follows. We can define an analogous 
PGL$(2,Q_p)$ valued scalar function $\mathfrak{g}(v)$, defined at each vertex, such that
\be
\mathfrak{g}(v) = \left(\begin{tabular}{cc} $p^n$ & $x^{(n)}$ \\
0 & 1
\end{tabular}\right),
\ee
where $(p^n, x^{(n)})$ is the coordinate of the vertex $v$. 
Then a Wilson line connecting two vertices $v_1, v_2$ is given by
\be
 \mathfrak{W}_{\textrm{padic}}(v_1\to v_2) = \mathfrak{g}(v_1)^{-1} \mathfrak{g}(v_2).
\ee

It evaluates to 
\begin{equation}
\mathfrak{W}_{\textrm{padic}}(v_1\to v_2)=\left(
\begin{array}{cc}
 p^{n_2-n_1}  & p^{-n_1}(x_2^{(n_2)}-x_1^{(n_1)}) \\
 0 & 1 \\
\end{array}
\right),
\end{equation}
with
\begin{eqnarray}
\label{v1}
  v_1=(p^{n_1},x_1^{(n_1)}),\\
  \label{v2}
  v_2=(p^{n_2},x_2^{(n_2)}).
\end{eqnarray}

This result would agree with the AdS Wilson lines given in (\ref{eq:wilads3}) if we blindly make the replacement
 \be \label{eq:replace}
 e^{-r} \leftrightarrow p^n, \qquad \zeta \leftrightarrow x^{(n)}, \qquad  \mathfrak{W}(v_1\to v_2) \leftrightarrow  \mathfrak{W}_p(v_1\to v_2) 
\ee

The expectation value of a Wilson line $p+1$ point junction in this connection evaluated at representations corresponding to primaries of the p-adic CFT agrees precisely with the tensor network we proposed in \cite{Hung:2019zsk} that is mentioned in the previous section. 
We would like to see if such a correspondence continues to hold when we consider the BTZ connection.

\subsection{Transplanting the BTZ connection to the Bruhat-Tits tree}
Consider the AdS metric (\ref{metric}) where $L(\zeta)$ does not vanish. 
Consider the specific case where $L(\zeta)$ is a constant, which reduces to the metric of the BTZ black hole. We
can obtain the SL$(2,\mathbb{C})$ connection that follows from this metric, and subsequently a Wilson line connecting two points $v_1, v_2$ as follows:
\begin{equation}
\label{wv1v2}
\mathfrak{W}(v_1\to v_2)=\left(
\begin{array}{cc}
 e^{-\delta r} \cosh \left(\frac{\delta\zeta \sqrt{L}}{\sqrt{k}}\right) & \frac{e^{r_1 } \sqrt{k} \sinh \left(\frac{\delta\zeta \sqrt{L}}{\sqrt{k}}\right)}{\sqrt{L}} \\
 \frac{e^{-r_2 } \sqrt{L} \sinh \left(\frac{\delta\zeta \sqrt{L}}{\sqrt{k}}\right)}{\sqrt{k}} & \cosh \left(\frac{\delta \zeta \sqrt{L}}{\sqrt{k}}\right) \\
\end{array}
\right)e^{\delta r/2},
\end{equation}

Of course (\ref{metric}) corresponds to pure AdS space for all $L(\zeta)$. On the other hand they would be related to the standard Poincare frame differently as we pick different $L(\zeta)$. Locally, the coordinates can be taken as $r,\zeta$ for all these different frames. 

Equipped with (\ref{eq:replace}), we are now ready to transplant the BTZ black hole connection to the Bruhat-Tits tree. As in the AdS case, the local coordinates describe a particular branch $k$  in equation (\ref{eq:bhcoord1}), and looks like $(p^{h+l}, y^{(h+l)})$.   They are related to $(r,\phi,\tau) $ by (\ref{eq:rphitau}). 
We now transplant (\ref{wv1v2}) and treat it as a PGL$(2,Q_p)$ Wilson lines, and replace the coordinates by local coordinates on the Bruhat-Tits tree. This gives 
\begin{equation}
\mathfrak{W}_{\textrm{padic}}(v_1\to v_2)=\left(
\begin{array}{cc}
 p^{l_2-l_1} \cosh_p \left(\delta x \sqrt{L}\right) & p^{-h-l_1} \sqrt{L}^{-1}\sinh_p \left(\delta x \sqrt{L}\right) \\
 p^{h+l_2} \sqrt{L} \sinh_p \left(\delta x \sqrt{L}\right) & \cosh_p \left(\delta x \sqrt{L}\right) \\
\end{array}
\right),
\end{equation}
where
\begin{eqnarray}
  v_1 &=& (r_1,x_1)=(p^{h+l_1},x_1), \nonumber \\
  v_2&=&(r_2,x_2)=(p^{h+l_2},x_2),   \nonumber  \\
  \delta x&=&x_2-x_1.    \label{eq:v1v2}
\end{eqnarray}

There is a mystery here. Exponentials appear in the connection. As we have emphasised many times in the previous section,
exponentials of p-adic numbers with finite norms could lead to divergence. As we are going to show in the next sub-section, if we take these exponentials and subsequently $\sinh$ and $\cosh$ to be those defined in (\ref{sinh}),(\ref{cosh}), we will recover complete agreement with the result of the tensor network.
 
One might wonder -- what happens had we insisted upon using the usual exponential in the computation of expectations of Wilson line networks? 
It turns out that it would lead to the {\it same} result  unless the Wilson lines touches the horizon, at which point all the network vanishes! The black hole horizon would appear so dark that branches emanating from the horizon essentially becomes disconnected. 
 The deformed definition preserves most of the results except to connect the branches together at the horizon exactly like what happens in the tensor network.  These will be made explicit in the next sub-section.

\subsection{Evaluating the Wilson line at fixed representations}
Having determined a connection, evaluating a Wilson line requires putting the connection into some given representations. 
It is too lengthy to give a complete review of constructing representations of PGL$(2,Q_p)$. For complete details, please refer to \cite{Hung:2018mcn}. 

Here we only collect the necessary expressions.
The basis we work with is based on the fact that functions of p-adic numbers form representations of the group PGL$(2,Q_p)$.  
For a given conformal primary of conformal $\Delta$, we can construct a basis
\be
|X; \Delta\rangle \equiv \tilde{\mathcal{O}}_\Delta (X) | 0\rangle, \qquad X \in Q_p
\ee
where 
\begin{eqnarray}
\tilde{\mathcal{O}}_\Delta (X) &=& \mathcal{N}(d,\Delta) \int_{Q_p} dY\, |X-Y|_q^{2\Delta - 2d} \mathcal{O}_\Delta(Y). \\
\mathcal{N}(d,\Delta) &=& \frac{    \zeta_p(2d-2\Delta) \zeta_p(2\Delta)}{\zeta_p(d-2\Delta) \zeta_p(2\Delta-d)} , \qquad
\zeta_p(s) = \frac{1}{1-p^{-s}}.
\end{eqnarray}
The state $|0\rangle$ is invariant under PGL$(2,Q_p)$ transformations, which is analogous to the vaccuum state of a CFT and $\tilde{\mathcal{O}}_{\Delta}(X)$ is referred to as a shadow operator in the CFT literature. The parameter $d$ is the dimension of the CFT, which we can also take to $d=1$ for simplicity. 

The corresponding bra basis states are given by
\be
\langle X;\Delta| \equiv \langle 0| \mathcal{O}_\Delta(X)
\ee
Under the action of a group element $g \in \textrm{PGL$(2,Q_p)$}$,
where
\be
g =\left( \begin{tabular}{cc}
$a$ & $b$\\
$c$&$d$ 
\end{tabular}\right), \qquad a,b,c,d \in Q_p,
\ee
 the states transform according to how the primary state transforms. i.e.
\be
\langle X;\Delta | \to \left\vert \frac{ad- bc}{(cX + d)^2}\right\vert^{\Delta}_p \bigg\langle\frac{aX + b}{cX+ d} ; \Delta \bigg\vert \qquad 
\ee

There is also a set of bra and ket states constructed from the primaries
\be
\langle \Delta \vert \equiv  \langle 0 | \mathcal{O}_\Delta(Z), \qquad | \Delta \rangle = \mathcal{O}_{\Delta}(0)| 0\rangle.
\ee
In usual CFT$_2$, $Z$ is often taken to be infinity. We showed in \cite{Hung:2018mcn} that the value of $Z$ drops out for any finite $Z$ when the end points of the Wilson lines are pushed to the asymptotic boundary. Therefore we can also think of $Z$ as a kind of gauge choice. 

The definition of these basis states ensure that
\be
\langle \Delta_i | X; \Delta_j\rangle = \delta_{ij} \delta(X-Z), \qquad \langle X;\Delta_i| \Delta_j\rangle = \frac{\delta_{ij}}{|X|_p^{2\Delta}}.
\ee

We can then evaluate a component of the open Wilson line operator using these identities above.
This gives
\begin{eqnarray}
 \langle\Delta|\mathfrak{W}_{\Delta}(v_1\to v_2)|\Delta\rangle &\equiv& \int_{Q_p} dX \langle \Delta | X;\Delta\rangle \mathfrak{W}(v_1\to v_2) \langle X;\Delta|\Delta\rangle
 \nonumber\\
  &=& \frac{p^{-(l_1+l_2)\Delta}}{\left|p^{l_2} \cosh_p \left(\delta x \sqrt{L}\right)Z+p^{-h} \sqrt{L}^{-1}\sinh_p \left(\delta x \sqrt{L}\right)\right|_p^{2\Delta}},  \label{eq:Wilsoncomp}
\end{eqnarray}
where we have used (\ref{identity}) and $v_1, v_2$ are still given by (\ref{eq:v1v2}).

In the following, we will choose $Z=0$. This choice ensures that at least for $L=0$,  (\ref{eq:Wilsoncomp}) reduces to the two point function that follows from the tensor network (\ref{OOvacuum}). The details of the derivation is relegated to the appendix.

\subsection{Open Wilson line expectation values in p-adic black hole}
After choosing $Z=0$, the expectation value of the open Wilson line in p-adic black hole connection becomes
\begin{eqnarray}
\nonumber
\langle\Delta|\mathfrak{W}_{\Delta}(v_1\to v_2)|\Delta\rangle
  &=& \frac{p^{-(l_1+l_2)\Delta}}{\left| p^{-h}\sqrt{L}^{-1}\sinh_p \left(\delta x \sqrt{L}\right)\right|_p^{2\Delta}}\\
  \label{OO5}
  &=&\frac{p^{-(2h+l_1+l_2)\Delta}}{\left| \sqrt{L}^{-1}\sinh_p \left(\delta x \sqrt{L}\right)\right|_p^{2\Delta}}. 
  \label{eq:Wilson2pt}
\end{eqnarray}

One can see that (\ref{OO5}) and (\ref{phiphi2}) are equal if we identify
 \be \label{eq:Lph}
 \sqrt L=\frac{1}{2p^h} .
 \ee  
Of course we know that (\ref{phiphi}) depends only on the number of edges separating the two vertices $v_1,v_2$, and so this result locally is really the same whether in the ``pure'' BT tree or the black hole background. Therefore locally, the black hole background is indeed identical to the ``pure'' BT tree. This is parallel to the fact that locally the BTZ black hole is the same as pure AdS space.

Let us also comment that one can see here that if the exponentials in the $\sinh, \cosh$ in (\ref{eq:Wilson2pt}) are replaced by the usual exponential, the denominator diverges whenever 
\be
|\delta x \sqrt{L}|_p  \ge 1,
\ee
and so these two point functions vanish -- signifying that the branches are disconnected as claimed in the previous section. The deformed exponential however allows the Wilson line computation to match with the tensor network exactly. i.e. the effect of local diffeomorphism leading to the change in the cutoff surface on the BT tree is reproduced by the ``deformed'' exponential in the connection, and not the usual exponential. 

When $p^h\to p^{-\infty}$, the points $v_1,v_2$ will live in the same branch since a single branch on the horizon already contains the entire Bruhat-Tits tree. Then,  $\phi_1=\phi_2$. Let's write $(r_1,x_1),(r_2,x_2)$ explicitly as
\begin{eqnarray}
  (r_1,x_1) &=& (p^{h+l_1},(p^h k_1,p^h\Theta(y_1))),\\
  (r_2,x_2) &=& (p^{h+l_2},(p^h k_2,p^h\Theta(y_2))),
\end{eqnarray}
with $k_1=k_2=k$ and
\begin{eqnarray}
  y_1 &=& \sum_{i=0}^{l_1-1}a^{(1)}_i p^i,\;\;\;(a^{(1)}_0\neq 0)\\
  y_2 &=& \sum_{i=0}^{l_2-1}a^{(2)}_i p^i.\;\;\;(a^{(2)}_0\neq 0)
\end{eqnarray}
Recall that our choice of $Z=0$ now matches the Wilson line expectation values in the bulk with the two-point correlation functions computed using the tensor network. Therefore we can rewrite the open Wilson line component as in (\ref{OO4}), and get
\begin{eqnarray}
\nonumber
  \langle\Delta|\mathfrak{W}_{\Delta}(v_1\to v_2)|\Delta\rangle
  &=& \frac{r_1^{-\Delta} r_2^{-\Delta}|p^{-h}|_p^{2\Delta}|\mathcal{E}_p(x_1/p^h)|_p^{\Delta}|\mathcal{E}_p(x_2/p^h)|_p^{\Delta}}
{|\mathcal{E}_p(x_1/p^h)-\mathcal{E}_p(x_2/p^h)|_p^{2\Delta}}\\
\nonumber
&=& \frac{p^{-(2h+l_1+l_2)\Delta }|p^{-h}|_p^{2\Delta}|p^k y_1|_p^\Delta|p^k y_2|_p^\Delta}{|p^k y_1-p^k y_2|_p^{2\Delta}}\\
\nonumber
&=&\frac{p^{-(l_1+l_2)\Delta }| y_1|_p^\Delta| y_2|_p^\Delta}{| y_1- y_2|_p^{2\Delta}}\\
&=&\frac{p^{-(l_1+l_2)\Delta }}{| y_1- y_2|_p^{2\Delta}}.
\label{eq:T0limit}
\end{eqnarray}
Here we have used $|y_1|_p=|y_2|_p=1$. And we notice that we actually recover the result (\ref{OOvacuum}) of the ``pure'' Bruhat Tits connection at $L=0$.

\section{Comments on p-adic black hole thermodynamics}
By making comparisons with real CFT$_2$, we have established that correlation functions computed in the black hole coordinates following from the tensor network describe some p-adic CFT at ``finite temperatures''.  

Separately, we worked with a p-adic BTZ black hole connection transplanted from the BTZ black hole connection in AdS$_3$ and computed the expectation value of an open Wilson line, and show that it agrees with the tensor network results with careful choice of the ambiguous parameter $Z$. It shows that indeed the p-adic BTZ connection encodes a finite temperature CFT exactly as in the AdS case. Now, we had the identification (\ref{eq:Lph}), and recall also (\ref{eq:Tph}), we land on the following relation
\be
\sqrt{L} = \frac{\mathcal{T}}{2}.
\ee
This is formally in complete agreement with the relation obtained from the BTZ black hole, although now $\sqrt{L}$ and $\mathcal{T}$ are both p-adic numbers.

Now then we can work backwards and ask for the horizon radius. Formally, in accordance with the BTZ black hole, the horizon radius $r_h$ is related to this parameter $\sqrt{L}$ by  \cite{Carlip:2005zn,Maldacena:1998bw}
\be
\sqrt{L} = \frac{r_h}{2 l}, \label{eq:Lrh}
\ee
 where we have already used the fact that we have set $l/(4G) = 1$.
 Using (\ref{eq:Lrh}), 
 \be
 |\mathcal{T}|_p=|p^{-h}|_p,
\ee
we finally have
 \be
 \left\vert \frac{r_h}{l}\right\vert_p = p^h. \label{eq:horizonrad}
 \ee
As promised, the transplanted BTZ connection does end up implying that the parameter we have introduced in the black hole coordinates back in (\ref{BHcoord}), is indeed giving us the radius of the black hole horizon. Let us also note that to make comparison with the radial coordinates in (\ref{BHcoord}), $p^h$ appears as a real number and so the comparison (\ref{eq:horizonrad}) only makes sense if we take the p-adic norm to map $r_h/l$ which is a p-adic number, to a real number.

When $|p^{-h}|_p=p^h\to p^{-\infty}$, we have $|\mathcal{T}|_p \to 0$ which is consistent with the limit $p^h\to p^{-\infty}$ considered in the computation (\ref{eq:T0limit}), which indeed recovers the original BT tree with vanishing $|\mathcal{T}|_p$.

Without an action, it is less clear how to define black hole entropy. On the other hand, the horizon radius $p^h$ is giving a scale to the density of branches emanated from the horizon. Recall the left most picture in Fig. \ref{cylinder}. 
Therefore, it is very intuitive to interpret this density of branches as being proportional to the entropy $S$, which we take as a measure of information and thus a real number here. 
If we take 
\be S \propto  p^h = |p^{-h}|_p, \implies S \sim |\mathcal{T}|_p \sim  |r_h|_p
\ee  
The p-adic BTZ black hole satisfies the area law just as its AdS sister. 

A determination of the constant of proportionality would require some extra input such as determining the Chern-Simons action defined on the tree exactly. This is beyond the scope of the current paper.

\section{Conclusion}

This is the second instalment of our series to bend the Bruhat-Tits tree -- that is to describe geometries that is deformed away from the ``pure'' Bruhat Tits tree. In our first instalment, the deformation is achieved by an RG flow that takes the p-adic CFT away from the fixed point. In the current instalment, we were inspired by the BTZ black hole, and considered transplanting the BTZ connection to a PGL$(2,Q_p)$ connection so that it describes a black hole in the Bruhat-Tits tree. Making use of the relation between these PGL$(2,Q_p)$ Wilson lines and the p-adic tensor network \cite{Hung:2019zsk}, we demonstrated not only that the expectation values of the Wilson lines can indeed be related to a p-adic CFT at finite temperatures, but more surprisingly that this black hole background can indeed be related to the ``pure'' BT tree by a local diffeomorphism, reflecting a non-trivial Weyl transformation performed in the p-adic CFT. 
Along the way,  the tree inspired us to define a p-adic ``deformed'' exponential function that shares many properties with the usual exponential that captures this local diffeomorphism. 

The results are quite surprising. It is known for a long time that the p-adic CFT does not appear to have descendents. But it is at the same time well known that non-trivial representations of the p-adic conformal group PGL$(2,Q_p)$ cannot be finite dimensional \cite{Ebert:2019src}. This paper is, to our knowledge, a first demonstration that it is possible to accommodate local conformal transformation in a way parallel to real 2d CFTs. 
This raises many questions, particularly whether Virasoro symmetry can in fact be defined in these p-adic CFTs after all. At first sight -- if it is possible to recover analogues of thermal correlators in the p-adic CFT, whether one could find cousins of the stress tensors that testify to some finite energy density at finite temperatures. Descendents and the stress tensor have eluded the p-adic CFT community for a long time. 
It would be very interesting as a first step to look for more examples of Weyl transformations connected to non-trivial diffeomorphism of the BT tree to gain insight of the problem. 

From the geometric perspective, we have got a glimpse of ``black hole'' geometries in the context of the Bruhat-Tits tree. The construction of the genus 1 tree via orbifolding was considered before \cite{Heydeman:2016ldy,Hung:2019zsk, Ebert:2019src}, but its connection to finite temperature physics is explored in depth only now. The relation between black hole temperature and horizon radius is carried through exactly to the BT tree. The black hole does appear to satisfy some notion of the area law, although a precise definition of the black hole entropy is still lacking, partially because much like anything p-adic, derivatives of the real valued partition function with respect to temperatures would not be well defined. One needs to be more creative to look for a proper measure of the entropy. 

We note that everything that is observed in this paper was inspired by the tensor network construction of the p-adic CFT partition function. Indeed it has guided us in recovering pieces of the p-adic AdS/CFT dictionary that is more universal than the tensor network reconstruction. The geometry considered in this paper is essentially locally a ``pure'' BT tree, and it would thus automatically satisfy the emergent Einstein equation discussed in \cite{paper1, paper2}. A very natural next step would thus be to synergise results of our two distinct approaches in bending the BT tree -- RG flow and diffeomorphism, to obtain more complicated scenarios that hopefully should help us understand the emergent Einstein equation we found in the prequel \cite{paper1,paper2}. 
Nonetheless, the present work should give further support that the tensor network does capture some very important essence of the p-adic AdS/CFT both qualitatively and quantitatively.  The tensor network should have more in supply in understanding new insights such as wormholes and islands \cite{Penington:2019npb,Almheiri:2019hni}. 
The tensor network should also give further insight in understanding complexities. Our tensor network has very precise relation with Wilson lines for example, which is strongly reminiscent of recent discussions of complexities based on a Wilson line network \cite{Chen:2020nlj}.

Last but not least, it may be confusing that we have, in our first and second instalment of the series, presented two apparently different interpretations of the tensor network. 
In the first paper, we have directly read off a metric from the tensor network, while in the current second instalment, we have made use of the Wilson line interpretation of the tensor network. One would have wondered whether the gauge connection and the edge distance read off from the tensor network can be related to each other in some way. The usual relation between vierbein and metric would not work, because we do not have an infinitesimal version of the gauge connection. What we have presented are two self-consistent complementary interpretations of the tensor network. It would be important to explore their connections further in the future. 

We hope that this success in the p-adic AdS/CFT reconstruction can ultimately be replicated to some extent in AdS/CFTs in general.

\textit{Acknowledgements.---} LYH acknowledges the support of NSFC (Grant
No. 11922502, 11875111) and the Shanghai Municipal Science and Technology Major Project
(Shanghai Grant No.2019SHZDZX01), and Perimeter Institute for hospitality as a part of the Emmy Noether Fellowship programme. 
Part of this work was instigated in KITP during the program qgravity20.  LC acknowledges support of NSFC (Grant No. 12047515). 
We thank  Bartek Czech, Ce Shen, Gabriel Wong, Qifeng Wu and Zhengcheng Gu for useful discussions and comments. 
We thank Si-nong Liu and Jiaqi Lou for collaboration on related projects.

\appendix

\section{Matching the tensor network bulk correlation functions to the Wilson line computation}

When the end points of the Wilson lines are not all pushed to the boundary, the components have dependence of
the parameter $Z$. We show below that there is a choice such that the result (\ref{eq:Wilsoncomp}) reduces to two point correlation functions in the bulk as well. 

\begin{figure}[htbp!]
\centering
\subfigure[]{
\label{position1}
\includegraphics[width=0.32\textwidth]{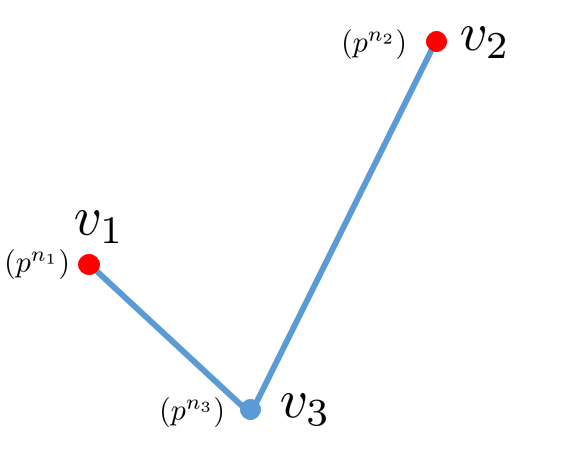}}
\subfigure[]{
\label{position2}
\includegraphics[width=0.32\textwidth]{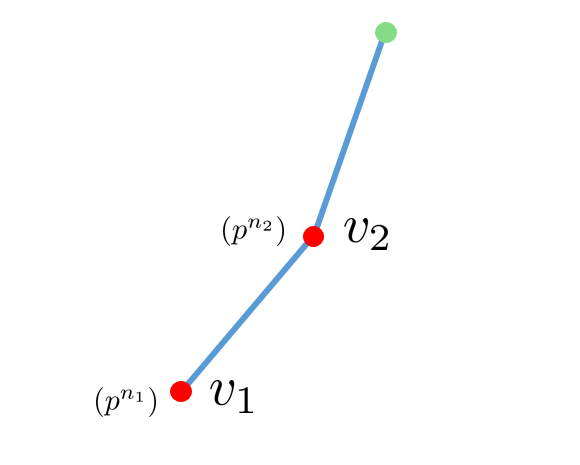}}
\subfigure[]{
\label{position3}
\includegraphics[width=0.32\textwidth]{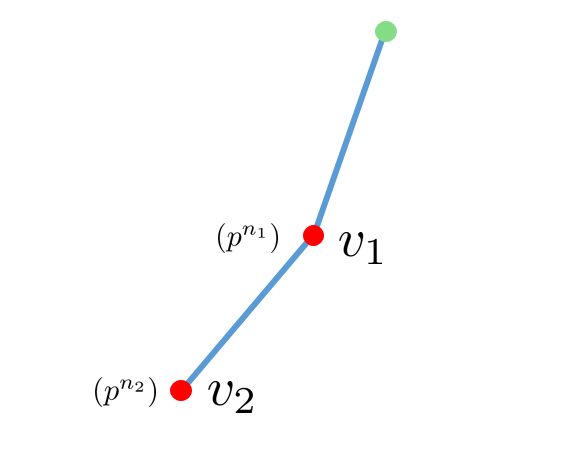}}
\caption{There are three types of relative positions. In (a), $|{\delta x}|_p=p^{-n_3}$. In (b), $|{\delta x}|_p=p^{-n_1}$. In (c), $|{\delta x}|_p=p^{-n_2}$. }
\label{position}
\end{figure}

When $L=0$ (``pure Bruhat-Tits connection''), we have
\begin{eqnarray}
  \langle\Delta|\mathfrak{W}_{\Delta}(v_1\to v_2)|\Delta\rangle
  &=& \frac{p^{-(n_1+n_2)\Delta}}{\left|p^{n_2} Z+ {\delta x}\right|_p^{2\Delta}}.
\end{eqnarray}
For p-adic number, when $|a|_p>|b|_p$, we have
\begin{eqnarray}
  |a+b|_p &=& |a|_p.
\end{eqnarray}
So when $|Z|_p < 1$, we have
\begin{eqnarray}
  \left|p^{n_2} Z+ {\delta x}\right|_p &=& \left| {\delta x}\right|_p,
\end{eqnarray}
since $|{\delta x}|_p\geq|p^{n_2}|_p>|p^{n_2}Z|_p$. 
Let us check what happens when we take $Z=0$.
\begin{eqnarray}
\langle\Delta|\mathfrak{W}_{\Delta}(v_1\to v_2)|\Delta\rangle
  &=& \frac{p^{-(n_1+n_2)\Delta}}{\left| {\delta x}\right|_p^{2\Delta}}.
\end{eqnarray}
It is explicitly symmetric in $v_1$ and $v_2$ as desired.
All the possible relative positions of $v_1$ and $v_2$ are shown in Fig\ref{position}.

In case (a), the two point function is
\begin{eqnarray}
\langle\Delta|\mathfrak{W}_{\Delta}(v_1\to v_2)|\Delta\rangle
  = \frac{p^{-(n_1+n_2)\Delta}}{p^{-2n_3\Delta}}=p^{-(n_1+n_2-2n_3)\Delta},
\end{eqnarray}
where $n_1+n_2-2n_3$ is the distance between $v_1$ and $v_2$.

In case (b), the two point function is
\begin{eqnarray}
 \langle\Delta|\mathfrak{W}_{\Delta}(v_1\to v_2)|\Delta\rangle  = \frac{p^{-(n_1+n_2)\Delta}}{p^{-2n_1\Delta}}=p^{-(n_2-n_1)\Delta},
\end{eqnarray}
where $n_2-n_1$ is the distance between $v_1$ and $v_2$.

In case (c), the two point function is
\begin{eqnarray}
\langle\Delta|\mathfrak{W}_{\Delta}(v_1\to v_2)|\Delta\rangle
  = \frac{p^{-(n_1+n_2)\Delta}}{p^{-2n_2\Delta}}=p^{-(n_1-n_2)\Delta},
\end{eqnarray}
where $n_1-n_2$ is the distance between $v_1$ and $v_2$.

In conclusion, we find that
\begin{eqnarray}
\label{OOvacuum}
  \langle\Delta|\mathfrak{W}_{\Delta}(v_1\to v_2)|\Delta\rangle
  = \frac{p^{-(n_1+n_2)\Delta}}{\left| \text{dx}\right|_p^{2\Delta}}=p^{-\Delta d(v_1,v_2)},
\end{eqnarray}
where $d(v_1,v_2)$ is the distance between $v_1$ and $v_2$. So after neglecting the term $p^{n_2} Z$, the two point function is the same as the two point function that would have followed from the tensor network discussed in section 2. Similarly, we can prove that after choosing $Z=0$, the three point function is the same as the one that follows from the tensor network.

\bibliographystyle{utphys}
 \bibliography{emergentE}

\end{document}